\documentclass[article,amsmath,amssymb,aps, twocolumn,nofootinbib,preprintnumbers,superscriptaddress]{revtex4-2}

\usepackage{natbib}
\usepackage{graphicx}
\usepackage{xcolor}
\usepackage{hyperref}
\hypersetup{colorlinks, allcolors=blue}
\usepackage{multirow}

\newcommand{\be}{\begin{equation}}
\newcommand{\ee}{\end{equation}}

\begin{document}
\bibliographystyle{apsrev4-1}

\title{Exploring the cosmological dark matter coincidence using infrared fixed points}

\author{Alexander C. Ritter}
 \email{rittera@student.unimelb.edu.au (corresponding author)}
\author{Raymond R. Volkas}
 \email{raymondv@unimelb.edu.au}
\affiliation{
ARC Centre of Excellence for Dark Matter Particle Physics,
School of Physics, \\
The University of Melbourne,
Victoria 3010, Australia
}

\begin{abstract}

The asymmetric dark matter (ADM) paradigm is motivated by the apparent coincidence between the cosmological mass densities of visible and dark matter, $\Omega_\mathrm{DM} \simeq 5\Omega_\mathrm{VM}$. However, most ADM models only relate the number densities of visible and dark matter, and do not motivate the similarity in their particle masses. One exception is a framework introduced by Bai and Schwaller, where the dark matter is a confined state of a dark QCD-like gauge group, and the confinement scales of visible and dark QCD are related by a dynamical mechanism utilising infrared fixed points of the two gauge couplings. We build upon this framework by properly implementing the dependence of the results on the initial conditions for the gauge couplings in the UV. We then reassess the ability of this framework to naturally explain the cosmological mass density coincidence, and find a reduced number of viable models. We identify features of the viable models that allow them to naturally relate the masses of the dark baryon and the proton while also avoiding collider constraints on the new particle content introduced.

\end{abstract}

\maketitle


\section{Introduction}

Determining the particle nature of dark matter (DM) is one of the deepest tasks facing particle physics today. This goal is hampered by the purely gravitational nature of the observational evidence for DM, which has allowed for a cornucopia of DM candidates to be proposed over a wide range of mass scales \cite{Feng:2010gw}. To focus our model-building pursuits, it is instructive to see what clues may lie in the existing astrophysical observations. 

One such result is the similarity between the present-day cosmological mass densities of dark and visible matter (VM), which we refer to as the \textit{cosmological coincidence} \cite{Planck:2018vyg}:
\be
\label{eqn:cosmo_coincidence}
    \Omega_{\mathrm{DM}} \simeq 5\Omega_\mathrm{VM},
\ee
where $\Omega_X$ is the mass density $\rho_X$ for species $X$ divided by the critical density $\rho_c$.

We take the cosmological coincidence to be a hint towards an underlying link between the origins of the abundances of VM and DM. Such a connection is not present in the majority of prominent DM candidates. Consider, for example, the WIMP: it is a GeV--TeV scale thermal relic species with an abundance generated through thermal freezeout. This stands in contrast to VM, whose number density is due to the baryon asymmetry of universe and whose mass arises from the confinement energy of QCD. With such distinct generation mechanisms, there is no \textit{a priori} reason why the cosmological mass densities of these species should fall in the same order of magnitude.

The main model-building paradigm that seeks to explain the cosmological coincidence problem is that of asymmetric dark matter (ADM) \cite{Petraki:2013wwa, Zurek:2013wia}. In this framework, the dark matter candidate is charged with a dark particle number that develops an asymmetry related to the visible baryon asymmetry. Models of ADM thus naturally generate similar number densities for VM and DM -- $n_{\mathrm{VM}} \sim n_{\mathrm{DM}}$. However, as a mass density is given by $\Omega_X = n_X m_X/\rho_c$, to satisfactorily explain the cosmological coincidence problem we must also motivate the similarity in the particle masses of VM and DM -- $m_{\mathrm{VM}} \sim m_{\mathrm{DM}}$. This problem is not addressed in the majority of the ADM literature, which mostly treats the DM mass as a free parameter and thus merely shifts the cosmological mass-density coincidence to a particle mass coincidence.

Our goal then is to construct a theory where DM and VM naturally have similar particle masses. By analogy with the proton, this leads us to consider DM candidates that are confined states of a dark, QCD-like gauge group \cite{Hodges:1993yb, Foot:2003jt, Foot:2004pq, Foot:2004pa, Ibe:2018juk, An:2009vq, Cui:2011wk, Lonsdale:2014yua, Farina:2015uea, Garcia:2015toa, Farina:2016ndq, Lonsdale:2017mzg, Lonsdale:2018xwd, Ibe:2019ena, Beauchesne:2020mih, Feng:2020urb, Ritter:2021hgu, Ibe:2021gil, Murgui:2021eqf, Bai:2013xga, Newstead:2014jva}, where some mechanism is present to relate the confinement scale $\Lambda_{\mathrm{dQCD}}$ of this gauge group to that of visible QCD, $\Lambda_{\mathrm{QCD}}$.

Previous efforts to relate the visible and dark confinements scales have followed one of two general directions:

\begin{enumerate}
    \item introduce a symmetry between the gauge groups. This generally leads to models of mirror matter, where the mirror symmetry is either exact \cite{Hodges:1993yb, Foot:2003jt, Foot:2004pq, Foot:2004pa, Ibe:2018juk} or judiciously broken \cite{An:2009vq, Cui:2011wk, Lonsdale:2014yua, Farina:2015uea, Garcia:2015toa, Farina:2016ndq, Lonsdale:2017mzg, Lonsdale:2018xwd, Ibe:2019ena, Beauchesne:2020mih,Feng:2020urb, Ritter:2021hgu, Ibe:2021gil} (For other models of mirror matter, see Refs.~\cite{Lee:1956qn, Kobzarev:1966qya, Berezhiani:1996sz, Blinnikov:1982eh, Foot:1991bp, Foot:1991py, Ignatiev:2003js, Foot:2004wz, Berezhiani:2000gw, Chacko:2005pe}).
    \item introduce new field content so that the running gauge couplings $\alpha_s$ and $\alpha_d$ approach infrared fixed points with similar magnitudes. This has not been widely discussed in the literature, with the original idea introduced by Bai and Schwaller \cite{Bai:2013xga} and expanded upon by Newstead and TerBeek \cite{Newstead:2014jva}.
\end{enumerate}

In this paper we focus on the latter approach, developing upon the framework introduced by Bai and Schwaller by properly implementing the dependence of the confinement scale $\Lambda_\mathrm{dQCD}$ on the initial values of the running gauge couplings in the UV, $\alpha_s^\mathrm{UV}$ and $\alpha_d^\mathrm{UV}$. We then ask how readily this approach generates confinement scales of the same order of magnitude, and so reassess the validity of this framework as a natural explanation of the cosmological coincidence.

Compared to the work of Bai and Schwaller, we find that in our analysis there are a smaller number of models that naturally generate similar confinement scales for visible and dark QCD, where models in this framework are defined by the new field content we introduce. This reduced set of models is due to our definition of `naturalness' now being more stringent, as it must take into account the dependence of the dark confinement scale on the initial gauge coupling values in the UV.

In general, we find that models whose infrared fixed points have small values for the gauge couplings are better at generically generating similar visible and dark confinement scales. However, we also find that these models generally require the mass scale of the new particle content to be sub-TeV for most selections of the initial couplings in the UV, and thus are subject to strong collider constraints. 

Looking for models where the new physics mass scale is on the order of a few TeV, we do find a number of such models that can fairly generically generate related visible and dark confinement scales. We identify these models as the most promising candidates within this framework for naturally explaining the cosmological coincidence problem.

The paper is organised at follows: in Section~\ref{sec:bs} we describe the dark QCD framework of Bai and Schwaller. In Section~\ref{sec:tc} we describe the threshold corrections to the running of the gauge couplings as implemented by Newstead and TerBeek. In Section~\ref{sec:uv} we analyse the effect of the values of the gauge couplings in the UV on $\Lambda_{\mathrm{dQCD}}$. In Section~\ref{sec:cc} we discuss the ability of this framework to explain the cosmological coincidence, before presenting our results in Section~\ref{sec:results} and concluding in Section~\ref{sec:conclusions}.


\section{The Bai-Schwaller framework}\label{sec:bs}

The scheme of Bai and Schwaller \cite{Bai:2013xga} utilises a dark QCD-like gauge group $SU(N_d)_{\textrm{dQCD}}$, along with a selection of new particle content charged under $SU(3)_{\textrm{QCD}}\times SU(N_d)_{\textrm{dQCD}}$. As in the original paper, we only consider $N_d = 3$, and limit the new field content to fermions and scalars in the fundamental representations of either one or both of the QCD gauge groups. The multiplicities of the new particles are given in Table~\ref{tab:particle_content}, along with their masses. In addition to the given multiplicities, we also define $n_{f_d} = n_{f_{d,h}} + n_{f_{d,l}}$ as the total number of dark fermion species that are fundamentals of $SU(3)_{\textrm{dQCD}}$, and $n_{f_c} = n_{f_{c,h}} + 6$ as the multiplicity of visible fermions that are fundamentals of $SU(3)_{\textrm{QCD}}$ (including the 6 SM quarks).

The majority of the new field content is, for simplicity, taken to exist at a common heavy mass scale $M$, except for the $n_{f_{d,l}}$ light dark fermions. These latter particles, which we refer to as `dark quarks', have masses much lighter than the dark confinement scale $\Lambda_\mathrm{dQCD}$ and are confined into the dark baryons that serve as the DM candidate. 

\begin{table}
\begin{tabular}{c|c|c|c}
\hline \hline
Field                    & $SU(3)_{\mathrm{QCD}}\times SU(3)_{\mathrm{dQCD}}$ & Mass                        & Multiplicity  \\ \hline \hline
\multirow{4}{*}{Fermion} & $(\mathbf{3},\mathbf{1})$                          & $M$                         & $n_{f_{c,h}}$ \\ \cline{2-4} 
                         & \multirow{2}{*}{$(\mathbf{1},\mathbf{3})$}         & $< \Lambda_{\mathrm{dQCD}}$ & $n_{f_{d,l}}$ \\ \cline{3-4} 
                         &                                                    & $M$                         & $n_{f_{d,h}}$ \\ \cline{2-4} 
                         & $(\mathbf{3},\mathbf{3})$                          & $M$                         & $n_{f_j}$     \\ \hline \hline
\multirow{3}{*}{Scalar}  & $(\mathbf{3},\mathbf{1})$                          & $M$                         & $n_{s_c}$     \\ \cline{2-4} 
                         & $(\mathbf{1},\mathbf{3})$                          & $M$                         & $n_{s_d}$     \\ \cline{2-4} 
                         & $(\mathbf{3},\mathbf{3})$                          & $M$                         & $n_{s_j}$     \\ \hline \hline
\end{tabular}
\caption{The new particle content in a model. The given multiplicities are for Dirac fermions and complex scalars. A subscript containing $l$ ($h$) indicates a multiplicity for light (heavy) particles in cases where this is a relevant distinction to make. \label{tab:particle_content}}
\end{table}

At two-loop level, the $\beta$-functions for $g_c$ and $g_d$ are coupled, thanks to the presence of the `joint' fields charged under both $SU(3)_{\mathrm{QCD}}$ and $SU(3)_{\mathrm{dQCD}}$. The two-loop $\beta$-function for $g_c$, $\beta_c(g_c, g_d) = \frac{dg_c}{d(\log(\mu))}$, is given by
\begin{equation}\label{eqn:beta_fn}
\begin{split}
    &\beta_c = \frac{g_c^3}{16\pi^2}\left[\frac{2}{3}\left(n_{f_c} + 3n_{f_j}\right)+\frac{1}{6}\left(n_{s_c}+3n_{s_j}\right)-11\right]\\
    &+\frac{g_c^5}{(16\pi^2)^2}\bigg[\frac{38}{3}\left(n_{f_c}+3n_{f_j}\right) + \frac{11}{3}\left(n_{s_c}+3n_{s_j}\right) - 102\bigg]\\
    &+\frac{g_c^3g_d^2}{(16\pi^2)^2}\left[8n_{f_j} + 8n_{s_j}\right],
\end{split}
\end{equation}
and the $\beta$-function for $g_d$, $\beta_d(g_c, g_d) \equiv \frac{dg_d}{d(\log(\mu))}$, is obtained by exchanging the indices $c \leftrightarrow d$ \cite{Jones:1981we}. Note that these $\beta$-functions are given for $N_c = N_d = 3$; the $\beta$-functions for general $N_c$ and $N_d$ can be found in Ref.~\cite{Bai:2013xga}.

Depending on the particle content, there may be nontrivial couplings $\alpha_s^*$ and $\alpha_d^*$ for which both $\beta$-functions are zero, where $\alpha_s = g_c^2/4\pi$, $\alpha_d = g_d^2/4\pi$, and
\be
\label{eqn:fixed_point}
    \beta_c(g_c^*, g_d^*) = \beta_d(g_c^*, g_d^*) = 0.
\ee
The couplings $\alpha_s^*$ and $\alpha_d^*$ denote an infrared fixed point (IRFP) of the renormalization group running similar to a Banks-Zaks fixed point for a single gauge coupling \cite{Banks:1981nn}. The couplings at the IRFP only depend on the multiplicities of the new particle content, $(n_{f_{c,h}}, n_{f_{d,l}}, n_{f_{d,h}}, n_{f_j}, n_{s_c}, n_{s_d}, n_{s_j})$; we refer to a given selection of multiplicities as a `model'.

So, for a given model, if there exists a nontrivial perturbative IRFP, the coupled gauge couplings evolve toward $\alpha_s^*$ and $\alpha_d^*$ regardless of the initial values for the gauge couplings in the UV. This fixed point feature is broken when the new heavy fields decouple at the mass scale $M$; below this scale, the gauge couplings run independently until they become non-perturbative at the confinement scales $\Lambda_{\mathrm{QCD}}$ and $\Lambda_{\mathrm{dQCD}}$.

The process to calculate $\Lambda_{\mathrm{dQCD}}$ for given model is then, ideally, as follows:
\begin{itemize}
    \item Calculate the IRFP gauge couplings $\alpha_s^*$ and $\alpha_d^*$.
    \item Determine $M$ by running the visible coupling up from the measured value $\alpha_s(M_Z) = $ 0.11729 \cite{ParticleDataGroup:2020ssz} to $\alpha_s(M) = \alpha_s^*$.
    \item Evolve the dark coupling down from $\alpha_d(M) = \alpha_d^*$ until the scale $\Lambda_{\mathrm{dQCD}}$ at which it becomes large enough to trigger confinement. As a rough perturbative condition, the Cornwall-Jackiw-Tomboulis bound $\alpha_s > \pi/4$ for $N_d = 3$ \cite{Cornwall:1974vz} is used to define the dark confinement scale through $\alpha_d(\Lambda_{\mathrm{dQCD}}) = \pi/4$.
\end{itemize}

So, for a given model with a nontrivial IRFP, the dark confinement scale $\Lambda_\mathrm{dQCD}$ can be uniquely determined. The argument is then that a model with related IRFP values $\alpha_s^*$ and $\alpha_d^*$ could lead to compatible confinement scales for visible and dark QCD of a similar order of magnitude.

However, this framework relies upon a number of simplifying assumptions that should be investigated in greater detail. The first is that the heavy fields decouple at the mass scale $M$ without any threshold corrections, which was addressed in Ref.~\cite{Newstead:2014jva}; we summarise their approach in the next section. The second key assumption is that the couplings run all the way to their IRFP values by the decoupling scale $M$ regardless of the initial coupling values in the UV. However, this is not true in general, and in a later section we analyse how the value of $\Lambda_{\mathrm{dQCD}}$ depends on the UV coupling values $\alpha_s^{\mathrm{UV}}$ and $\alpha_d^{\mathrm{UV}}$.


\section{Threshold corrections}\label{sec:tc}

In MS-like mass-independent renormalization schemes, the Appelquist-Carazzone decoupling theorem \cite{Appelquist:1974tg} does not apply in its naive form; heavy fields can continue to influence coupling constants and $\beta$-functions at energy scales below their masses. In these schemes there also arises the related issue of large logarithms when working at energy scales much smaller than the mass $M$ of the heavy fields.

To properly account for these problems, the decoupling is treated explicitly by constructing an effective field theory in which the heavy fields have been integrated out. Consistency is ensured by matching the full theory onto the effective field theory at a matching scale $\mu_0$, where $\mu_0 \sim M$ to avoid large logarithms. This procedure leads to a `consistency condition' relating the coupling constants in the full theory with the coupling constants in the effective field theory \cite{Weinberg:1980wa, Ovrut:1980dg},
\begin{align}
    &\alpha_s^{\textrm{EFT}}(\mu_0) = \zeta_c^2(\mu_0, \alpha_s(\mu_0))\alpha_s(\mu_0), \label{eqn:cons_cond_s}\\
    &\alpha_d^{\textrm{EFT}}(\mu_0) = \zeta_d^2(\mu_0, \alpha_s(\mu_0))\alpha_d(\mu_0). \label{eqn:cons_cond_d}
\end{align}

The decoupling functions $\zeta_c^2$ and $\zeta_d^2$ have been determined to three- and four-loop order in Refs.~\cite{Chetyrkin:1997un, Chetyrkin:2005ia} for the case of integrating out one heavy quark from QCD with $n_f$ flavours. Since we are working with two-loop $\beta$-functions, we only require the one-loop decoupling functions; in Ref.~\cite{Newstead:2014jva}, the results of Ref.~\cite{Chetyrkin:1997un} were adapted for integrating out the full selection of heavy field content at mass $M$, with the one-loop decoupling functions given by
\begin{align}
    &\zeta_c^2(\mu, \alpha(\mu)) = 1 - \frac{\alpha_s(\mu)}{6\pi}\tilde{n}_c\log\left(\frac{\mu^2}{M^2}\right),\\
    &\zeta_d^2(\mu, \alpha(\mu)) = 1 - \frac{\alpha_d(\mu)}{6\pi}\tilde{n}_d\log\left(\frac{\mu^2}{M^2}\right),
\end{align}
where the coefficients $\tilde{n}_c$ and $\tilde{n}_d$ account for the degrees of freedom of the relevant heavy fields,
\begin{align}
    &\tilde{n}_c = n_{f_{c,h}} + 3 n_{f_j} + \frac{1}{4}\left(n_{s_c} + 3 n_{s_j}\right),\\
    &\tilde{n}_d = n_{f_{d,h}} + 3 n_{f_j} + \frac{1}{4}\left(n_{s_d} + 3 n_{s_j}\right).
\end{align}

We note that, for a given $M$, the values of the couplings at low energies now depend on the choice of decoupling scale $\mu_0$; this renormalization scale dependence is a non-physical artifact of our perturbative analysis being truncated at finite loop order. As a result, the procedure of Bai and Schwaller (matching $\alpha_s(M_Z)$ to its experimental value) no longer uniquely determines the value of $M$ when we incorporate the threshold corrections to the coupling constants. Instead, we obtain a relationship between $M$ and $\mu_0$, and can solve for one of these values if we specify the other. 

In Ref.~\cite{Newstead:2014jva}, values of $\Lambda_{\mathrm{dQCD}}$ were determined for a given model and value of $M$. By using the consistency condition and again matching the strong coupling to its measured  value at $M_Z$, they solved for the decoupling scale $\mu_0$, and then ran the dark coupling until $\alpha_d(\Lambda_{\mathrm{dQCD}}) = \pi/4$. However, this process does not ensure that $\mu_0$ is on the order of $M$. We take a different approach to implementing threshold corrections, as is detailed in the following section.


\section{Dependence on initial UV couplings}\label{sec:uv}

\begin{figure}
    \centering
    \includegraphics[width=0.45\textwidth]{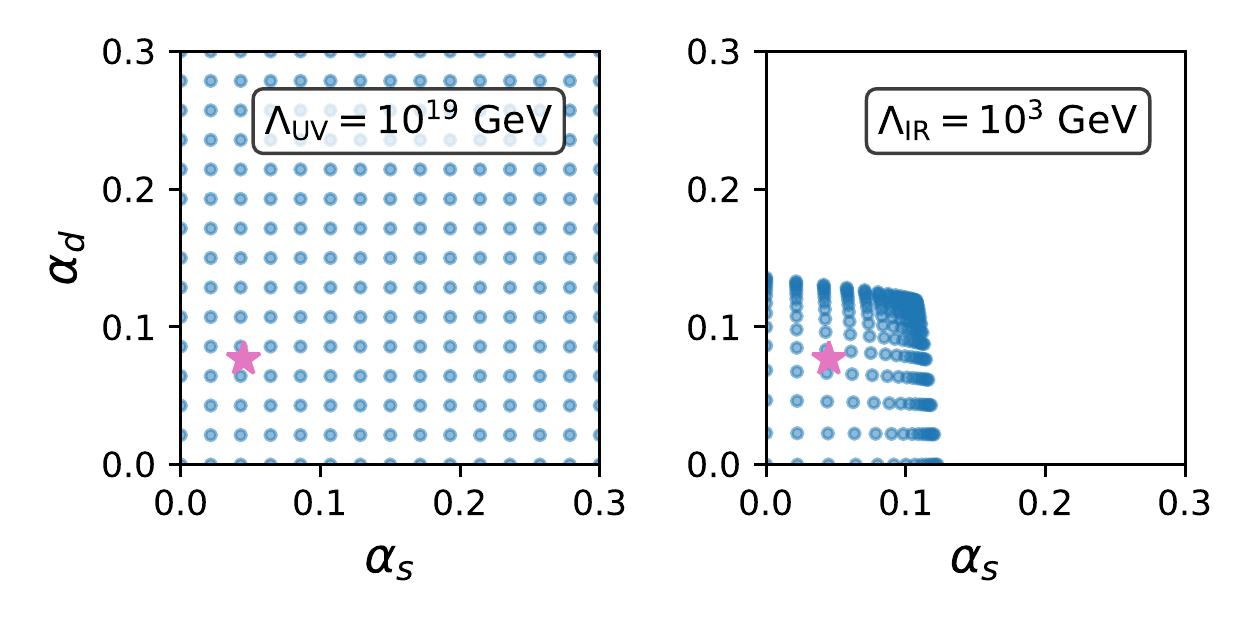}
    \caption{\textbf{Left:} A grid of values for the strong coupling constant $\alpha_s$ and the dark coupling constant $\alpha_d$ at an energy scale $\Lambda_\mathrm{UV} = 10^{19}$ GeV. The pink star shows the infrared fixed point $(\alpha_s^*, \alpha_d^*) = (0.045, 0.077)$ for the model with field multiplicities $(n_{f_{c,h}}, n_{f_{d,l}}, n_{f_{d,h}}, n_{f_j}, n_{s_c}, n_{s_d}, n_{s_j}) = (3, 3, 3, 3, 3, 2, 0)$. \textbf{Right:} The grid of values from the left panel has been evolved down from $10^{19}$ GeV to an energy scale $\Lambda_\mathrm{IR} = 10^{3}$ GeV under the RGEs of the given model. The pink star again indicates the IRFP of the model.}
    \label{fig:couplings_from_UV_to_IR}
\end{figure}

As mentioned earlier, the framework in the previous sections assumes that the couplings run to their IRFP values by the decoupling scale $\mu_0$ regardless of their initial values in the UV, $\alpha_s^\mathrm{UV}$ and $\alpha_d^\mathrm{UV}$. However, this is not the case, as illustrated in Fig.~\ref{fig:couplings_from_UV_to_IR} for the model $(n_{f_{c,h}}, n_{f_{d,l}}, n_{f_{d,h}}, n_{f_j}, n_{s_c}, n_{s_d}, n_{s_j}) = (3, 3, 3, 3, 3, 2, 0)$  with IRFP $(\alpha_s^*, \alpha_d^*) = (0.045, 0.077)$, where we show the running of a grid of couplings $(\alpha_s, \alpha_d)$ from the UV scale $\Lambda_\mathrm{UV} = 10^{19}$ GeV down to a typical infrared scale $\Lambda_\mathrm{IR} = 1$ TeV. While the couplings evolve towards their fixed point values, they do not precisely reach them, and so the low energy behaviour of the theory exhibits some dependence on the couplings in the UV.

\begin{figure}
    \centering
    \includegraphics[width=0.4\textwidth]{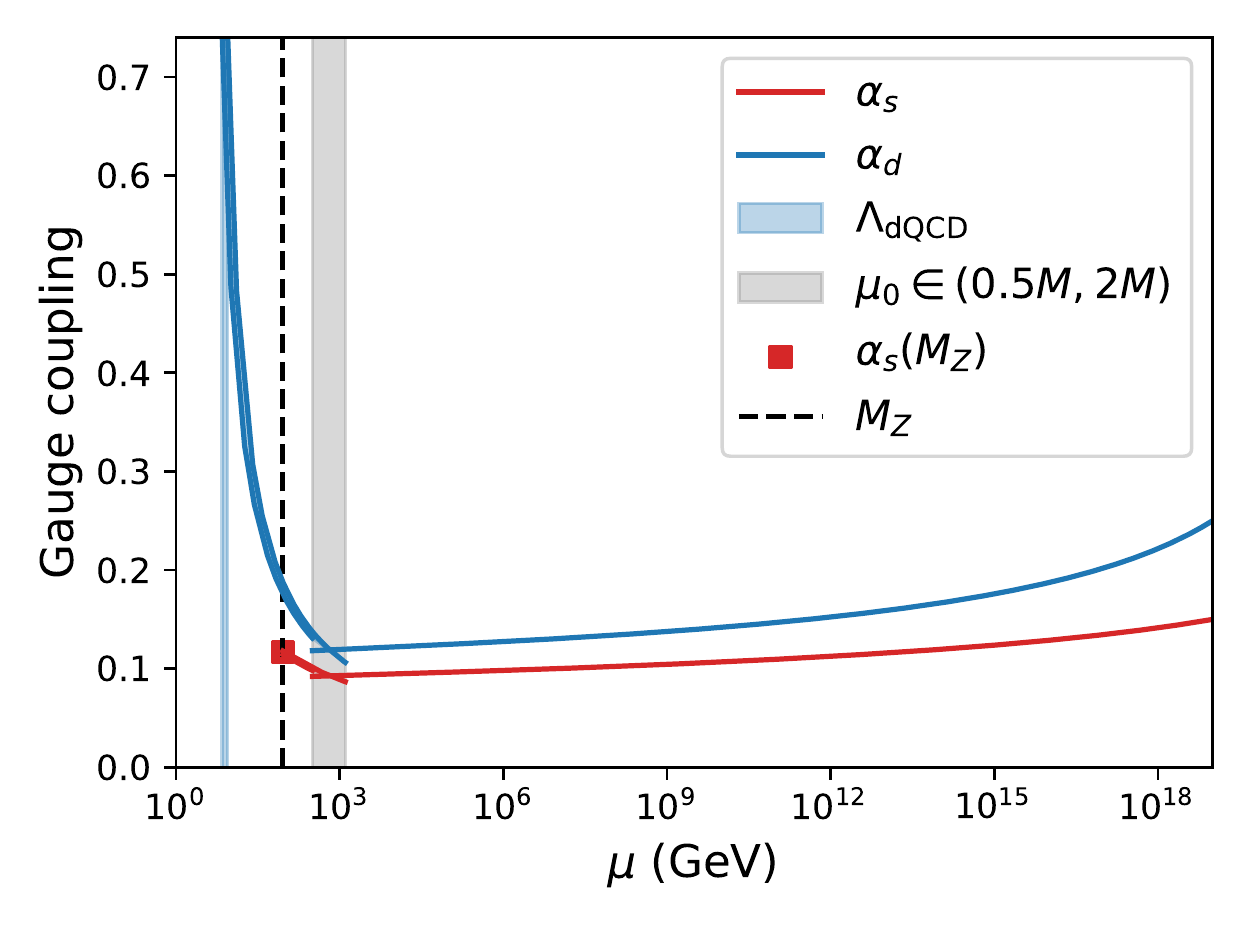}
    \includegraphics[width=0.4\textwidth]{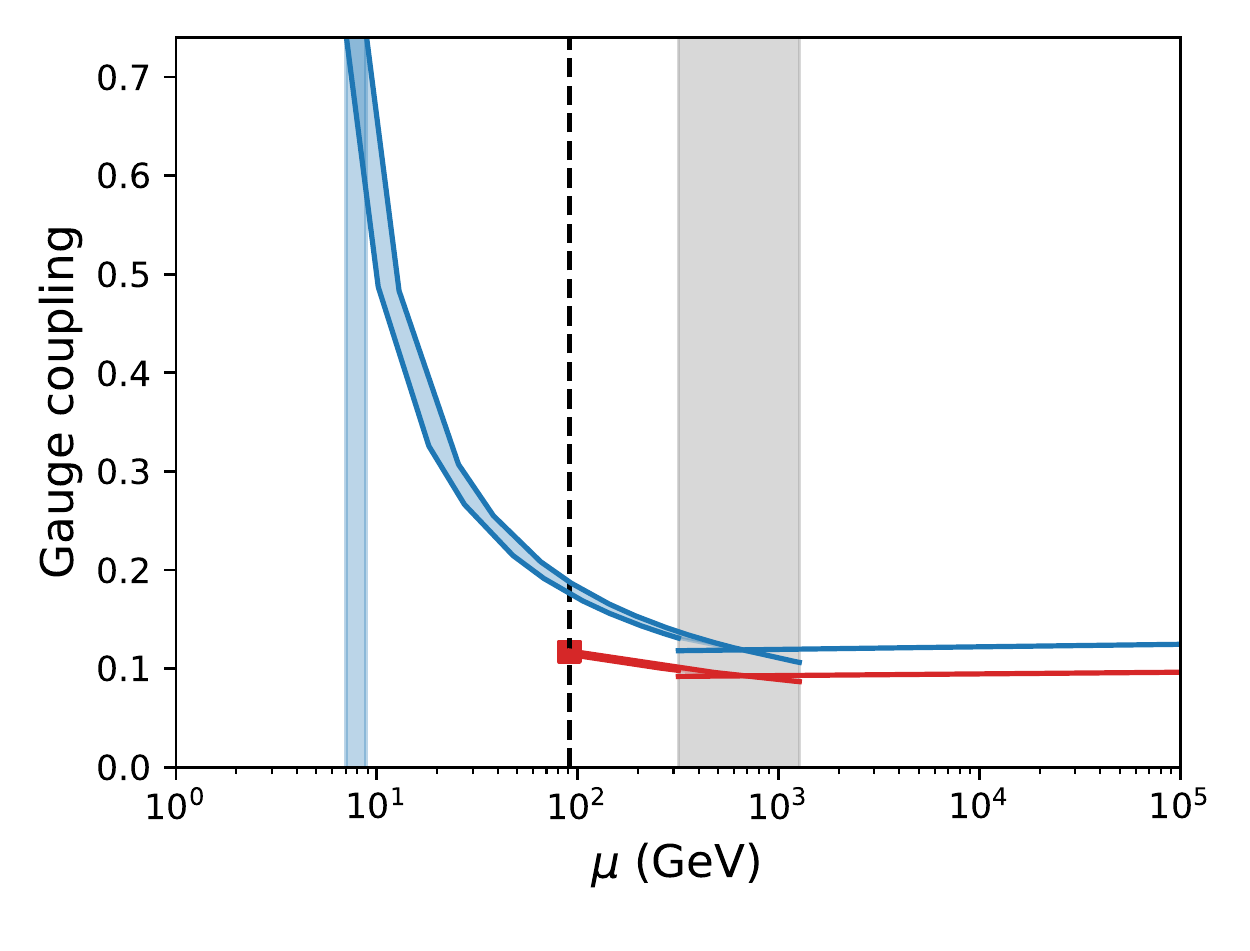}
    \caption{\textbf{Top:} The evolution of the QCD and dQCD coupling constants for the model $(n_{f_{c,h}}, n_{f_{d,l}}, n_{f_{d,h}}, n_{f_j}, n_{s_c}, n_{s_d}, n_{s_j}) = (3, 3, 3, 3, 3, 2, 0)$ with mass scale $M = 635$ GeV and initial UV couplings $(\alpha_s^\mathrm{UV}, \alpha_d^\mathrm{UV}) = (0.15, 0.25)$. The mass scale is chosen so that $\alpha_s$ obtains its measured value at $M_Z$, as indicated by the red square. The decoupling scale $\mu_0$ is shown by the grey shaded band, where we allow it to vary between $0.5M$ and $2M$ to account for the uncertainty introduced by threshold corrections. Below the decoupling scale, $\alpha_s$ and $\alpha_d$ are depicted as bands, due to the variance in the decoupling scale. The vertical blue band indicates the range of values for the dark confinement scale $\Lambda_\mathrm{dQCD}$, calculated as the energy scale where $\alpha_d = \pi/4$. \textbf{Bottom:} The same evolution, zoomed in to the energy range between 1 GeV and $10^5$ GeV to show the variance in the coupling constants below the decoupling scale more clearly.}
    \label{fig:sample_evolution}
\end{figure}

We now account for this dependence in our calculation of $\Lambda_\mathrm{dQCD}$. Consider the running of the couplings in this model for a given mass scale $M$ and initial UV couplings $(\alpha_s^\mathrm{UV}, \alpha_d^\mathrm{UV})$. The couplings evolve with the full coupled renormalization group equations until the scale $\mu_0$ at which we apply the matching conditions of Eqns.~\ref{eqn:cons_cond_s} and \ref{eqn:cons_cond_d}. Unlike the approach of Ref.~\cite{Newstead:2014jva}, we explicitly ensure that $\mu_0 \sim \mathcal{O}(M)$; following the approach of Refs.~\cite{Chetyrkin:2005ia,Davier:2005xq}, we vary $\mu_0$ between $0.5M$ and $2M$ and treat the variance in the low energy running of the couplings as a theoretical uncertainty due to working at finite loop-order.\footnote{Refs.~\cite{Chetyrkin:2005ia, Davier:2005xq} apply this approach to the running of $\alpha_s$ across quark thresholds in the SM between $m_\tau$ and $M_Z$. For a quark of mass $m_q$, they vary the decoupling scale between $0.7m_q$ and $3m_q$; however, since this process is intended to minimise large logarithms, we chose to vary the decoupling scale between multiplicatively symmetric values around $M$. Regardless of the specific choice made, note that this is merely a simple, arbitrary condition which roughly captures the uncertainty in the decoupling of the heavy fields. Indeed, other more sophisticated schemes have been developed to minimise the effect of unphysical renormalization scale dependence in threshold corrections (see for example Refs.~\cite{Brodsky:1982gc, Chishtie:2020cen, Stevenson:1981vj}) but these are beyond the scope of this analysis.} This then leads to an uncertainty in the value of $\Lambda_\mathrm{dQCD}$. This evolution is depicted in Fig.~\ref{fig:sample_evolution}, where the blue band shows the possible range of $\Lambda_\mathrm{dQCD}$ values for this given model, mass scale $M$ and initial UV couplings.

The process for calculating $\Lambda_\mathrm{dQCD}$ for a given model is now as follows:
\begin{itemize}
    \item We first specify the UV couplings $(\alpha_s^\mathrm{UV}, \alpha_d^\mathrm{UV})$; we then determine the required decoupling scale $\mu_0$ by solving Eqn.~\ref{eqn:cons_cond_s} where $\alpha_s^\mathrm{EFT}(\mu_0)$ is given by running the visible coupling up from $\alpha_s^\mathrm{EFT}(M_Z) = $ 0.11729 and $\alpha_s(\mu_0)$ is given by running $(\alpha_s^\mathrm{UV}, \alpha_d^\mathrm{UV}) \rightarrow (\alpha_s(\mu_0), \alpha_d(\mu_0))$.
    \item However, this does not uniquely specify $\mu_0$, as the decoupling function $\zeta_c^2$ depends on $M$. Since $\zeta_c^2$ depends on $M$ through the factor $\log(\mu^2/M^2)$, we can solve for $\mu_0$ (and also for $M$) by specifying a value for $\mu_0/M$. 
    \item To account for the uncertainty introduced by threshold corrections, we vary the value of $\mu_0/M$ between 0.5 and 2. This results in a range of values for $M$ for the given pair of initial UV couplings.
    \item We then calculate $\Lambda_\mathrm{dQCD}$ as usual by running $\alpha_d^\mathrm{EFT}(\mu_0)$ to $\alpha_d^\mathrm{EFT}(\Lambda_\mathrm{dQCD}) = \pi/4$. Since there is a range of possible values for $M$ (and thus $\mu_0$), we also obtain a range of values for $\Lambda_\mathrm{dQCD}$.
\end{itemize}

We note that this process will not work for all pairs of UV couplings, as we require the low energy running of $\alpha_s$ to match with experiment. This matching is performed by solving for the decoupling scale such that $\alpha_s$ reaches its measured value at $M_Z$, and for some initial UV couplings, there will be no value of the new physics mass scale $M$ between $M_\mathrm{Pl}$ and $M_Z$ for which this occurs.

\begin{figure}
    \centering
    \includegraphics[width=0.45\textwidth]{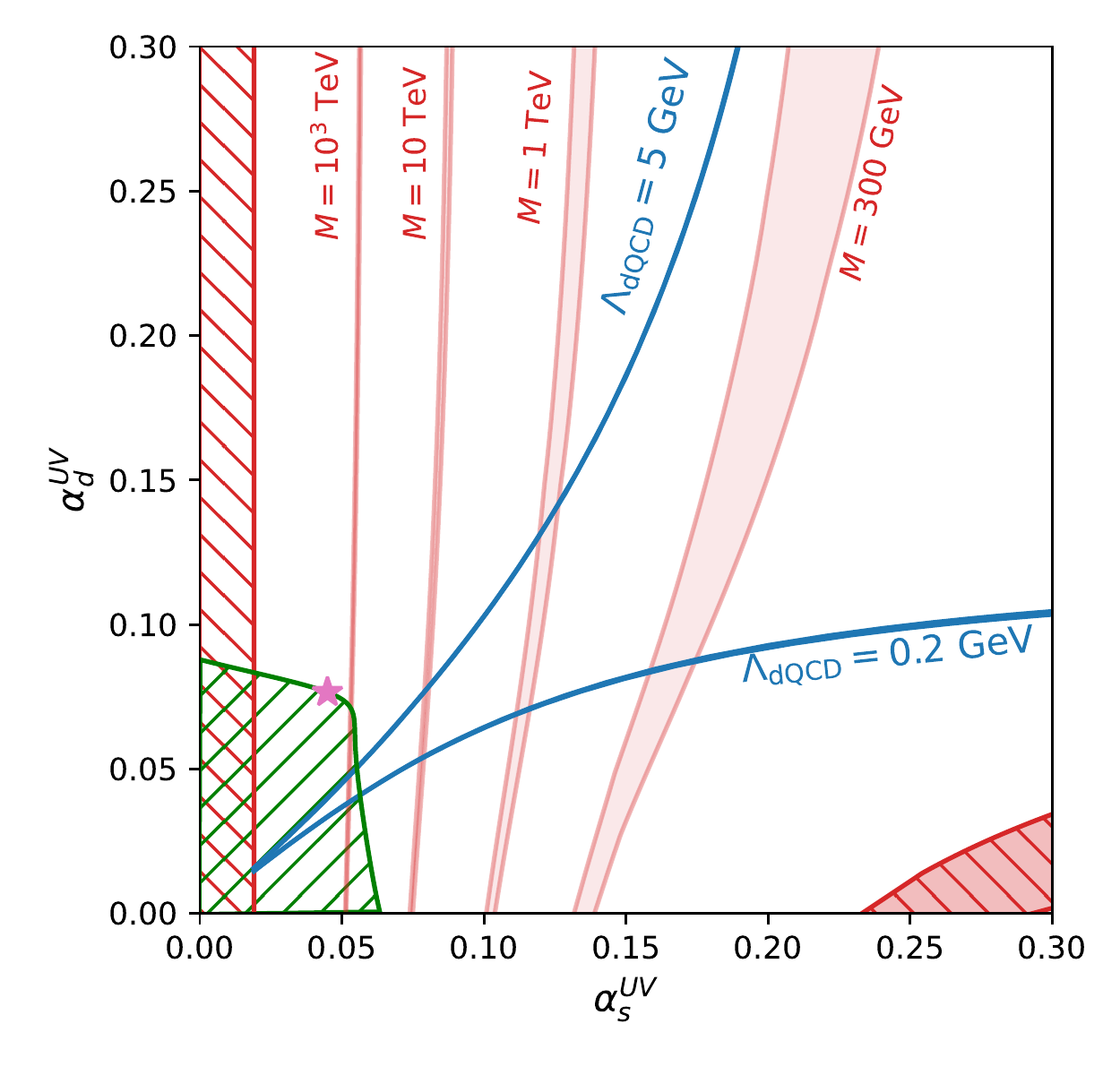}
    \caption{For the model $(n_{f_{c,h}}, n_{f_{d,l}}, n_{f_{d,h}}, n_{f_j}, n_{s_c}, n_{s_d}, n_{s_j}) = (3, 3, 3, 3, 3, 2, 0)$ we show contours for the dark confinement scale $\Lambda_\mathrm{dQCD}$ (blue) and for the new physics scale $M$ (red) on axes of the initial UV couplings $\alpha_s^\mathrm{UV}$ and $\alpha_d^\mathrm{UV}$. The width of these contours is due to the variance in the decoupling scale $\mu_0$ between $0.5M$ and $2M$ to account for the uncertainty introduced by threshold corrections. In the red hatched regions there are no allowed values of $M$ for which $\alpha_s$ has the correct running at low energy. The pink star shows the infrared fixed point of this model. The green hatched region is the `asymptotically free' region, where $SU(3)_\mathrm{QCD}$ and $SU(3)_\mathrm{dQCD}$ both exhibit asymptotic freedom.}
    \label{fig:sample_aUV_plot}
\end{figure}

With this process specified, we are interested in how $\Lambda_\mathrm{dQCD}$ depends on the initial UV couplings for a given model. As an example, in Fig.~\ref{fig:sample_aUV_plot} we plot contours of $\Lambda_\mathrm{dQCD}$ on axes of $\alpha_s^\mathrm{UV}$ against $\alpha_d^\mathrm{UV}$, where we work with perturbative couplings between 0 and 0.3 to ensure that we can trust our two-loop beta-function calculations\footnote{We discuss this choice for the range of initial UV couplings in greater detail in Sec.~\ref{sec:model_validity}.}. We see that $\Lambda_\mathrm{dQCD}$ is larger for small values of $\alpha_s^\mathrm{UV}$, and is smaller for small values of $\alpha_d^\mathrm{UV}$. Note that the contours are in fact shaded bands to account for the uncertainty in $\Lambda_\mathrm{dQCD}$; however, the threshold corrections do not have a drastic effect, as the bands are fairly narrow. 

In Fig.~\ref{fig:sample_aUV_plot} we also show contours for the required new physics mass scale $M$, which increases as $\alpha_s^\mathrm{UV}$ decreases. In the red hatched areas there are no allowed values of $M$ for which $\alpha_s$ has the correct running at low energy. 

\subsection{Asymptotic freedom}\label{sec:asymptotic_freedom}

In Fig.~\ref{fig:sample_aUV_plot} we have also included a star and a green hatched region. The star indicates the coupling values at the IRFP, and the green hatched region shows the `asymptotically free' region: that is, the region of the perturbative $(\alpha_s^\mathrm{UV}, \alpha_d^\mathrm{UV})$ parameter space for which $SU(3)_\mathrm{QCD}$ and $SU(3)_\mathrm{dQCD}$ both exhibit asymptotic freedom.

It is an interesting consequence of the existence of a nontrivial IRFP with coupled $\beta$-functions that the asymptotic freedom of the theory depends upon the values of the two gauge couplings. The intuition for this observation is as follows: as the energy scale approaches the IR, the gauge couplings approach their IRFP values, which we can also think of as points in the $(\alpha_s, \alpha_d)$ parameter space running toward the IRFP. So, as the energy scale approaches the UV, all points in the parameter space evolve away from the IRFP. There is a UV fixed point at the origin, $\alpha_s = \alpha_d = 0$; points that are roughly `closer' to the origin than the IRFP evolve towards this fixed point and thus both visible and dark QCD will exhibit asymptotic freedom. Those points `further away' from the origin than the IRFP instead diverge toward large couplings and do not exhibit asymptotic freedom. 

\begin{figure}
    \centering
    \includegraphics[width=0.4\textwidth]{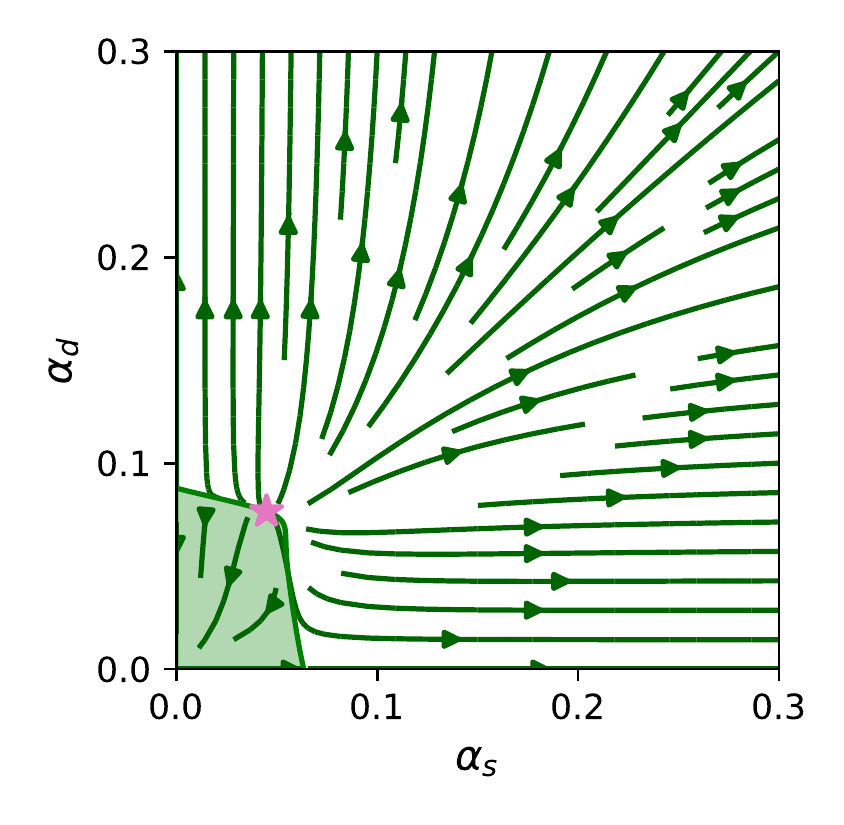}
    \caption{A streamline plot for the vector field defined by the coupled $\beta$-functions for $\alpha_s$ and $\alpha_d$ for the model $(n_{f_{c,h}}, n_{f_{d,l}}, n_{f_{d,h}}, n_{f_j}, n_{s_c}, n_{s_d}, n_{s_j}) = (3, 3, 3, 3, 3, 2, 0)$, where the green arrows indicate the change in the coupling constants with increasing energy scale $\mu$. The shaded green region is the `asymptotically free' region, corresponding to the green hatched region in Fig.~\ref{fig:sample_aUV_plot}. Within this region, both coupling constants approach zero as the energy scale increases. The pink star shows the infrared fixed point of this model.}
    \label{fig:af_streamlines}
\end{figure}

To see this region more concretely, we treat the $\beta$-functions as a vector field describing the flow of the couplings with increasing energy scale and plot its streamlines in Fig.~\ref{fig:af_streamlines}. We can clearly see the qualitative behaviour described in the previous paragraph, along with the edge of the asymptotically free region as shown in Fig.~\ref{fig:sample_aUV_plot}.

We note that although we have identified the region of $(\alpha_s, \alpha_d)$ parameter space for which the gauge groups are not asymptotically free, we do not discount this region. Since we work with $\alpha_s^\mathrm{UV}$ and $\alpha_d^\mathrm{UV}$ below 0.3, we only consider running in which the couplings are perturbative up to the Planck scale. Thus, dynamics at energies above the Planck scale may change the running again and regain asymptotic freedom or create a nonzero UV fixed point as per the ``asymptotic safety'' idea \cite{Weinberg:1976xy, Weinberg:1980gg}.


\section{Explaining the cosmological coincidence problem}\label{sec:cc}

Now that we have accounted for the dependence of $\Lambda_\mathrm{dQCD}$ on the initial UV couplings, in this section we reassess whether this framework offers a natural solution to the cosmological coincidence problem. 

To `solve' the cosmological coincidence problem, we mean that we wish our theory to naturally generate a mass density $\Omega_\mathrm{DM}$ on the order of $\Omega_\mathrm{VM}$. In this work we focus only on explaining the similarity between the particle masses, $m_\mathrm{VM} \sim m_\mathrm{DM}$, and do not specify the ADM mechanism through which the number densities are related.\footnote{For an example of an ADM mechanism that can be implemented in this framework, see the leptogenesis model of Ref.~\cite{Bai:2013xga}.} We assume that the ADM mechanism generates similar number densities $n_\mathrm{VM} \sim n_\mathrm{DM}$, and also assume that the relationship between the baryon mass and confinement scale is the same for both visible and dark QCD. Then, to satisfy Eqn.~\ref{eqn:cosmo_coincidence}, we need a dark confinement scale $\Lambda_\mathrm{dQCD} \sim 1$ GeV. We let this vary by a factor of 5 to approximately account for different dynamical completions that would be expected to produce a range of reasonable values for the $n_\mathrm{VM}/n_\mathrm{DM}$ ratio. So, we define a `valid' value of $\Lambda_\mathrm{dQCD}$ to be between 0.2 GeV and 5 GeV.

We now need to determine how `natural' this framework is; that is, how readily it generates valid values for $\Lambda_\mathrm{dQCD}$. We consider the theory to provide a natural explanation of the coincidence problem if one obtains a valid $\Lambda_\mathrm{dQCD}$ value for a large proportion of the available parameter space, as we do not wish for the cosmological coincidence to arise from any fine-tuning of the parameters. 

Of course, an ambiguity arises in the notion of `available' parameter space. The parameters we vary are the multiplicities of the field content and the initial values of the couplings in the UV; clearly, we must place some restrictions on these parameters when searching for valid $\Lambda_\mathrm{dQCD}$ values, both to provide a finite space in which to look and to avoid considering unphysical theories. So, the question we ask of the theory is as follows: \emph{given a certain set of assumptions, is it a surprise that we obtain a valid value for the dark confinement scale?}

In this section we do not argue for a unique set of such assumptions; rather, we discuss a number of assumptions one could make, and give freedom to the reader to place their own set of requirements on the theory when assessing its validity. We then present a set of results given only a minimal set of assumptions, and highlight features of the theory that affect how readily it obtains a valid dark confinement scale.

\subsection{Assessing the validity of a given model}\label{sec:model_validity}

We start this discussion by considering a given model, where we fix the multiplicities of the new fields as given in Table~\ref{tab:particle_content}. This specifies the IRFP and the beta-functions, but leaves us free to vary the initial UV couplings $\alpha_s^\mathrm{UV}$ and $\alpha_d^\mathrm{UV}$. The question then is what assumptions we place on the selection of these initial UV couplings; in particular, what range of values do we select our couplings from, and how do we randomly select coupling values from this range?

To answer this question requires knowledge of the inaccessible dynamics above the Planck scale that generate the values of these parameters at the UV scale. We must remain agnostic to these dynamics, and so the simplest assumption we apply is that the gauge couplings are perturbative at the UV scale -- that is, both $\alpha_s^\mathrm{UV}$ and $\alpha_d^\mathrm{UV}$ lie between 0 and 1. 

However, as seen in Figs.~\ref{fig:couplings_from_UV_to_IR} and \ref{fig:sample_aUV_plot}, we have chosen to work with initial UV couplings that lie between 0 and 0.3. This is done to ensure the perturbative validity of our two-loop $\beta$-function calculations. From Eqn.~\ref{eqn:beta_fn}, and recalling that $\alpha=g^2/4\pi$, we see that the individual one- and two-loop non-gluonic terms that contribute to the $\beta$-functions -- which we call $\beta^{(1)}$ and $\beta^{(2)}$, respectively -- are of the form
\be
\beta^{(1)} \sim \frac{\alpha^2n}{2\pi}, \quad \beta^{(2)} \sim 20\frac{\alpha^3n}{8\pi^2},
\ee
where $n$ is a multiplicity, and the factor of ~20 accounts for the relative size of the numerical coefficients for the one- and two-loop terms, due to various group-theoretic factors. Then, to ensure that the individual two-loop terms are smaller than the individual one-loop terms implies the rough condition $\alpha_s, \alpha_d < 0.3$. From now on, we refer to gauge couplings smaller than 0.3 as `perturbative' couplings, in the stricter sense that both the couplings themselves are perturbative, and that the $\beta$-functions governing their RGE running admit a perturbative analysis.

Other restrictions we could place on $\alpha_s^\mathrm{UV}$ and $\alpha_d^\mathrm{UV}$ depend upon how strongly we require a given model to replicate our universe. For example, one could discount points in the red hatched regions of Fig.~\ref{fig:sample_aUV_plot} where there is no value of the new physics scale $M$ for which the low energy running of $\alpha_s$ matches with experiment. However, one could argue that a given model is equally theoretically valid with any perturbative values for the initial UV couplings; the fact that the model fails this phenomenological test for certain initial UV couplings is just one way in which it is unable to obtain a valid value for the dark confinement scale. Put another way, the difference is between asking the following questions:
\begin{enumerate}
    \item ``Given that our model has perturbative values for $\alpha_s^\mathrm{UV}$ and $\alpha_d^\mathrm{UV}$, how likely is it that we replicate the low-energy running of $\alpha_s$ and also obtain a valid value for the dark confinement scale?"
    \item ``Given that our model has perturbative values for $\alpha_s^\mathrm{UV}$ and $\alpha_d^\mathrm{UV}$ and replicates the low-energy running of $\alpha_s$, how likely is it that we obtain a valid value for the dark confinement scale?"
\end{enumerate}
For the results we present in Sec.~\ref{sec:results}, we take the first of these approaches when analysing a given model.

Another possible assumption of this type is to only consider UV couplings for which the required value of $M$ is on the order of a TeV or greater, as new sub-TeV colour-charged particles would be produced readily at colliders. 

For a given model, one could also choose to only work with UV couplings for which the model is truly asymptotically free. However -- as discussed in Sec.~\ref{sec:asymptotic_freedom} -- given that we only consider $\alpha_s^\mathrm{UV}$ and $\alpha_d^\mathrm{UV}$ with values smaller than 0.3, then there will be no Landau poles below the Planck scale. Above the Planck scale, unknown physics could lead these models to regain asymptotic freedom or achieve asymptotic safety, so it is valid to consider such points.

The last point to consider here is the process by which we randomly select values for $\alpha_s^\mathrm{UV}$ and $\alpha_d^\mathrm{UV}$ from a specified range. This again requires knowledge of physics above the Planck scale; given our lack of knowledge, we make the simple assumption that the couplings are selected uniformly randomly between 0 and 0.3 for each coupling. 

Given the simplest of these assumptions, we define the `validity fraction' $\epsilon_v$ to be the fraction of the UV coupling parameter space -- $0 < \{\alpha_s^\mathrm{UV}, \alpha_d^\mathrm{UV}\} < 0.3$ -- that lies between the 0.2 GeV and 5 GeV contours for $\Lambda_\mathrm{dQCD}$. This is roughly the likelihood that a given model will obtain a valid value of $\Lambda_\mathrm{dQCD}$ given a random choice of perturbative initial UV couplings. As an approximate condition, we consider models with large values of $\epsilon_v$ to be those that explain the cosmological coincidence problem. The definition of a `large' value of $\epsilon_v$ is, of course, subjective to some extent, and we shall see what constitutes the largest possible values of $\epsilon_v$ when we present results in Sec.~\ref{sec:results}.

\subsection{Assessing the validity of the framework}

Now that we have discussed the assumptions one could make to assess the validity of a given model, we can look at how readily the overall framework can serve as an explanation of the coincidence problem. That is, within the space of all possible models (as defined by the multiplicities of the new fields we introduce), how many of them provide viable explanations of the cosmological coincidence problem? To answer this question, we must look at the assumptions we choose to restrict this space of all possible models.

In Ref.~\cite{Newstead:2014jva}, the space of all possible models is set by calculating a `maximum multiplicity' for each new field; these are determined by the requirement that the one-loop terms for $\beta_c$ and $\beta_d$ in Eqn.~\ref{eqn:beta_fn} remain negative.\footnote{In Ref.~\cite{Newstead:2014jva}, this condition is chosen to ensure that the gauge groups are QCD-like. In our work, we often work with gauge couplings where the models are not asymptotically free, so it is less obvious that we should also require negative one-loop terms in our $\beta$-functions. However, the existence of an IRFP ensures that there is an asymptotically free region surrounding the origin -- as shown in Fig.~\ref{fig:af_streamlines} -- and to have asymptotic freedom for $\alpha_s$ and $\alpha_s$ close to zero requires that the one-loop terms in the $\beta$-functions are negative.} For each new field, its maximum multiplicity is calculated assuming that all other new field multiplicities are zero; the resulting values are given in Table~\ref{tab:max_mult}.

\begin{table}
\begin{tabular}{ccccccc} \hline \hline
Multiplicity  & $n_{f_{c,h}}$ & $n_{f_d}$ & $n_{f_j}$ & $n_{s_c}$ & $n_{s_d}$ & $n_{s_j}$ \\ \hline
Maximum value & 10 & 16 & 3 & 42 & 66 & 14 \\ \hline \hline
\end{tabular}
\caption{The maximum multiplicities for each new field. These are set by the requirement that the one-loop terms for the $\beta$-functions are negative, which is necessary for the existence of an infrared fixed point. \label{tab:max_mult}}
\end{table}

Within this total space of possible models, we can apply further restrictions on the set of models we consider. The first, of course, is to consider only those models with a perturbative infrared fixed point. We could also consider only those models whose IRFP satisfies $\alpha_s^\star < \alpha_s(M_Z)$ and $\alpha_d^\star < \pi/4$. Before accounting for the dependence of the dark confinement scale on the initial UV couplings, this was a necessary condition to apply to a given model to even obtain a dark confinement scale; since the couplings were assumed to take their IRFP values at the decoupling scale, these values must be lower than the values they need to reach at low energy after decoupling. However, with our alteration to the framework, there will be initial UV coupling values for which a dark confinement scale can be calculated, even for models whose IRFP values $\alpha_s^\star$ and $\alpha_d^\star$ are larger than $\alpha_s(M_Z)$ and $\pi/4$, respectively; thus, we are not required to apply this restriction to the models we consider.

\section{Results}\label{sec:results}

In this section we present results using a minimal set of the assumptions discussed in the previous section. We focus on finding models that can viably explain the coincidence problem, and identifying the features of these models that allow them to do so \footnote{While this section focuses on identifying viable models within our expanded framework, in Appendix~\ref{sec:bs_models} we present results for the seven benchmark models from the original work of Bai and Schwaller; this serves as a point of comparison between the new and old analyses.}. To identify these viable models, we use the validity fraction $\epsilon_v$ as defined in Sec.~\ref{sec:model_validity}.

The entire space of possible models, as limited by the maximum field multiplicities of Table~\ref{tab:max_mult}, is too large to search through efficiently. We begin this section by instead looking in detail at a limited set of models - those that have at most three copies of each new field, as specified by the field identifications in Table~\ref{tab:particle_content}. We also require that $n_{f_{d,l}} \geq 1$, so that there is at least one light dark quark species to confine into the dark baryons. The choice of at most three copies of each field is only to make the set of models more amenable to study; however, these models also would present an easier task if one wished to do further model-building and detailed phenomenology with one of them in particular.

\subsection{Analysing models with $n_i \leq 3$}\label{sec:models_3}

\begin{figure}
    \centering
    \includegraphics[width=0.45\textwidth]{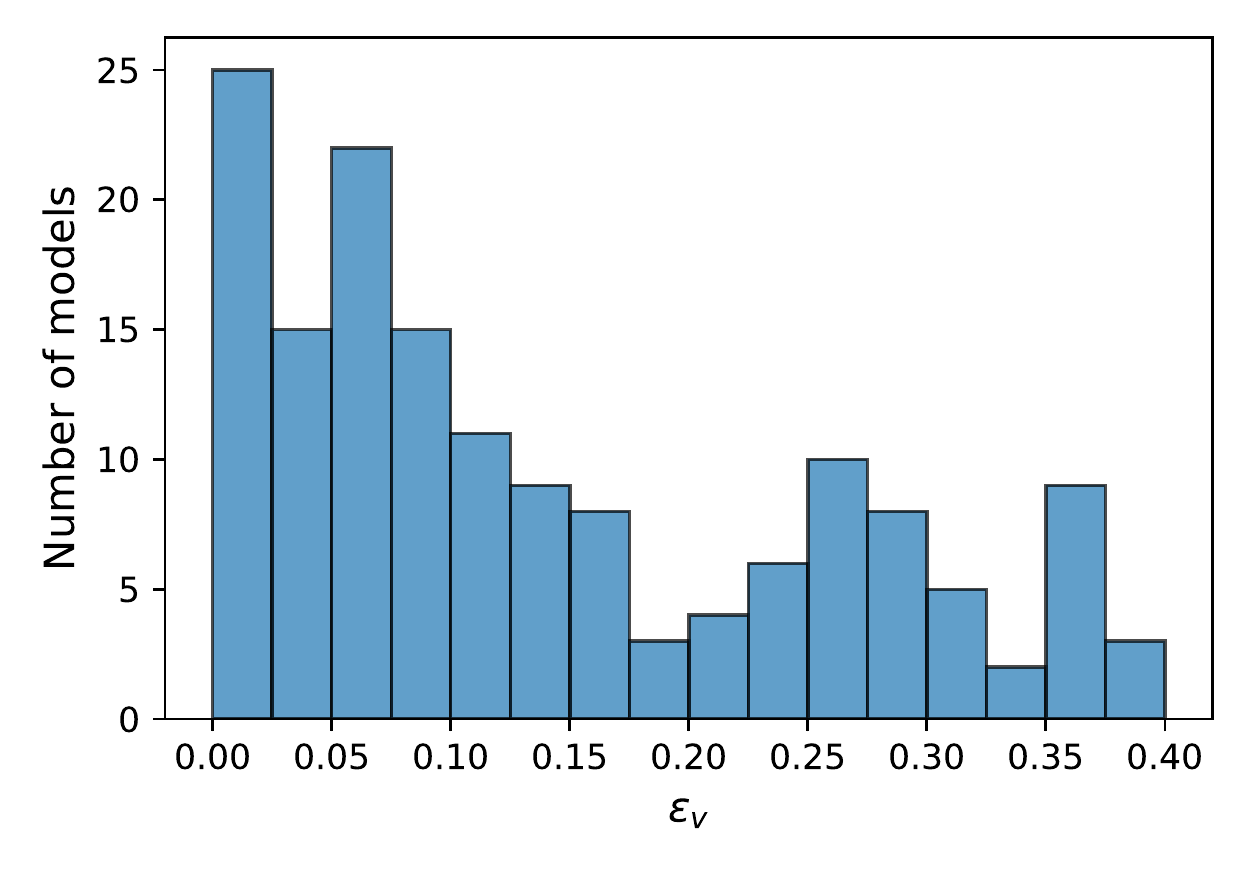}
    \caption{A histogram of the validity fractions for the set of 155 models that have at most three of each new field and a perturbative IRFP.}
    \label{fig:epsilon_v}
\end{figure}

There are 12,288 total models of this type. Of these, 1354 have an infrared fixed point and 155 have an IRFP for which both couplings are perturbative; recall that we consider a gauge coupling to be perturbative if it is smaller than 0.3, as discussed in Sec.~\ref{sec:model_validity}. The values of $\epsilon_v$ for this set of 155 models are shown in Fig.~\ref{fig:epsilon_v}. These are no models for which $\epsilon_v$ is greater than 0.4, and there is a tendency towards smaller values of $\epsilon_v$. 

\begin{figure*}
    \centering
    \includegraphics[width=0.95\textwidth]{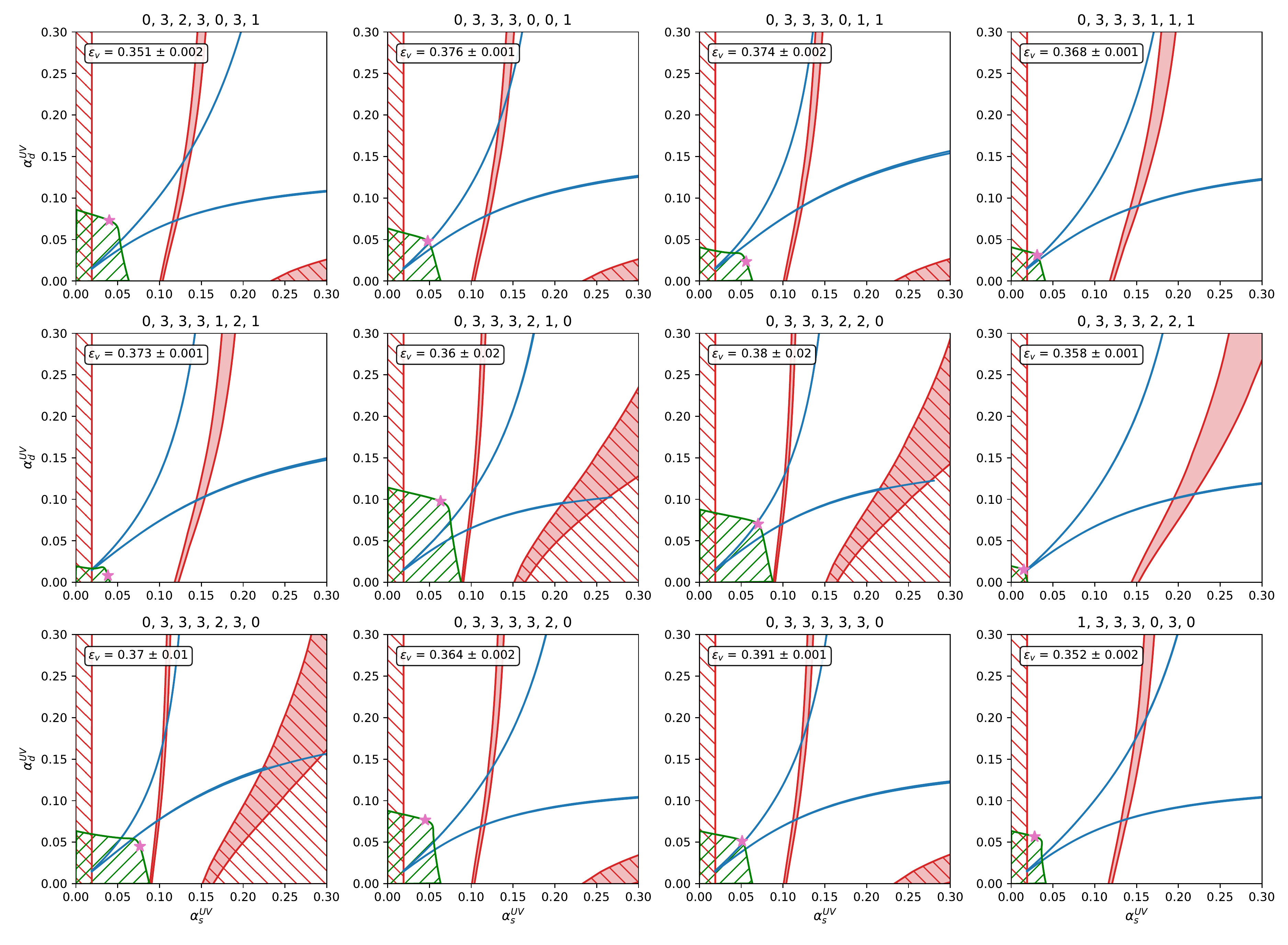}
    \caption{Each subplot shows the results for a given model on axes of the initial UV couplings, where the plot features are similar to those in Fig.~\ref{fig:sample_aUV_plot}. In blue we show the contours for $\Lambda_\mathrm{dQCD} =$ 0.2 GeV (right-most contour) and 5 GeV (left-most contour); the validity fraction $\epsilon_v$ is calculated as the proportion of the parameter space between these two contours. In red we show the contour for $M$ = 1 TeV, and note that $M > 1$ TeV to the left of the red contour; we also show the red hatched regions where there are no valid values for $M$. The green hatched region indicates the `asymptotically free' region, and the pink star shows the IRFP of the model. The 12 models shown are those with at most 3 of each new field, a perturbative IRFP, and a validity fraction $\epsilon_v > 0.35$. The title of each subplot gives the field multiplicities of the given model: $n_{f_{c,h}}, n_{f_{d,l}}, n_{f_{d,h}}, n_{f_j}, n_{s_c}, n_{s_d}, n_{s_j}$.}
    \label{fig:viable_models}
\end{figure*}

We consider models with larger values of $\epsilon_v$ to be those that are best able to explain the cosmological coincidence problem; we thus are interested in the features of the models with the largest values of $\epsilon_v$, and so in Fig.~\ref{fig:viable_models} we show the 12 models for which $\epsilon_v > 0.35$. Roughly, these are the models for which there is at least a 1 in 3 chance that a randomly chosen pair of perturbative initial UV couplings will generate similar values of the visible and dark confinement scales.

\begin{figure}
    \centering
    \includegraphics[width=0.45\textwidth]{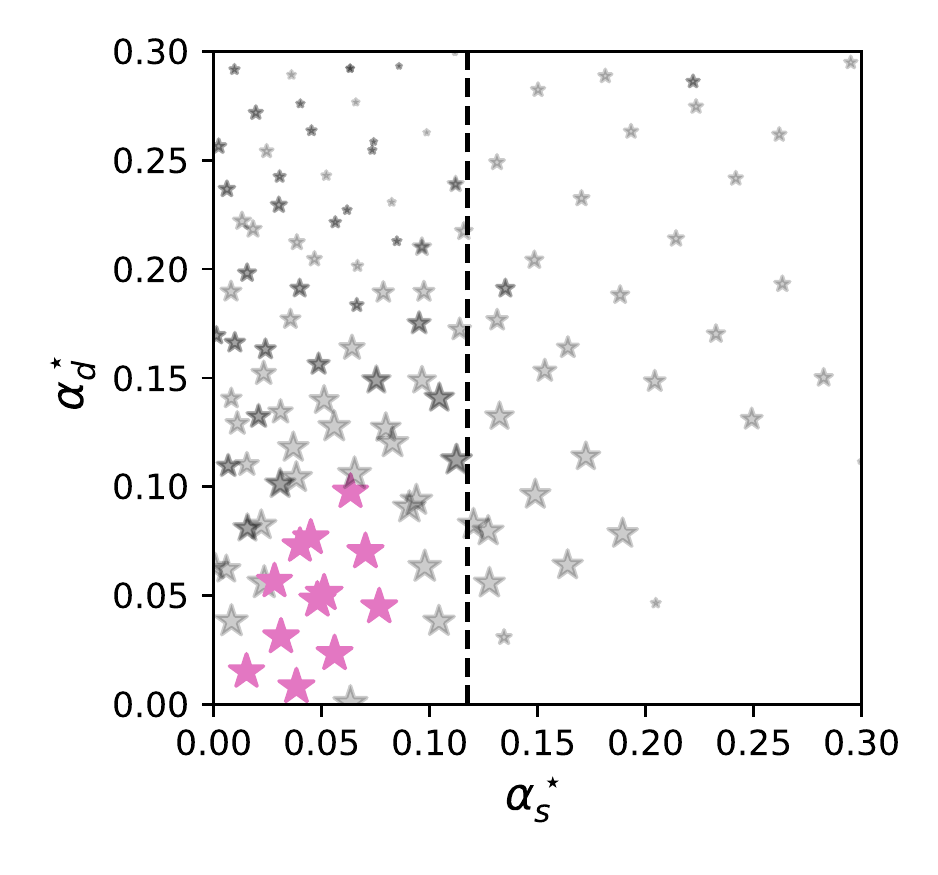}
    \caption{Each star shows the infrared fixed point of a model, sized by the validity fraction $\epsilon_v$ for that model. The set of models shown in this plot are the 155 models with at most 3 of each new field and a perturbative infrared fixed point. The pink stars correspond to the 12 models with $\epsilon_v > 0.35$ that are shown in Fig.~\ref{fig:viable_models}. The black dashed line is at $\alpha_s(M_Z) = $ 0.11729.}
    \label{fig:viable_irfps}
\end{figure}

Looking at the $\Lambda_\mathrm{dQCD}$ contours for these 12 models, we can first see why there may be no models for which $\epsilon_v$ is greater than 0.4. When the initial UV couplings are too dissimilar, we do not obtain a dark confinement scale similar to the visible confinement scale; for smaller $\alpha_s^\mathrm{UV}$ and larger $\alpha_d^\mathrm{UV}$, $\Lambda_\mathrm{dQCD}$ is too large, and for smaller $\alpha_d^\mathrm{UV}$ and larger $\alpha_s^\mathrm{UV}$, $\Lambda_\mathrm{dQCD}$ is too small. This makes sense -- to obtain similar confinement scales for the two gauge groups, their couplings must be similar at the decoupling scale, and so cannot be too different in the UV.

The next observation we make is that all 12 models have small fixed points, with both $\alpha_s^\star$ and $\alpha_d^\star$ less than 0.1. This correlation between smaller IRFP values and larger validity fractions is clear in Fig.~\ref{fig:viable_irfps}, where we plot the IRFPs for the set of 155 models, sized by the value of $\epsilon_v$. The 12 models with $\epsilon_v > 0.35$ are shown in pink, and cluster in a region centred on $\alpha_s^\mathrm{UV}, \alpha_d^\mathrm{UV} \sim 0.05$.

\begin{figure}
    \centering
    \includegraphics[width=0.47\textwidth]{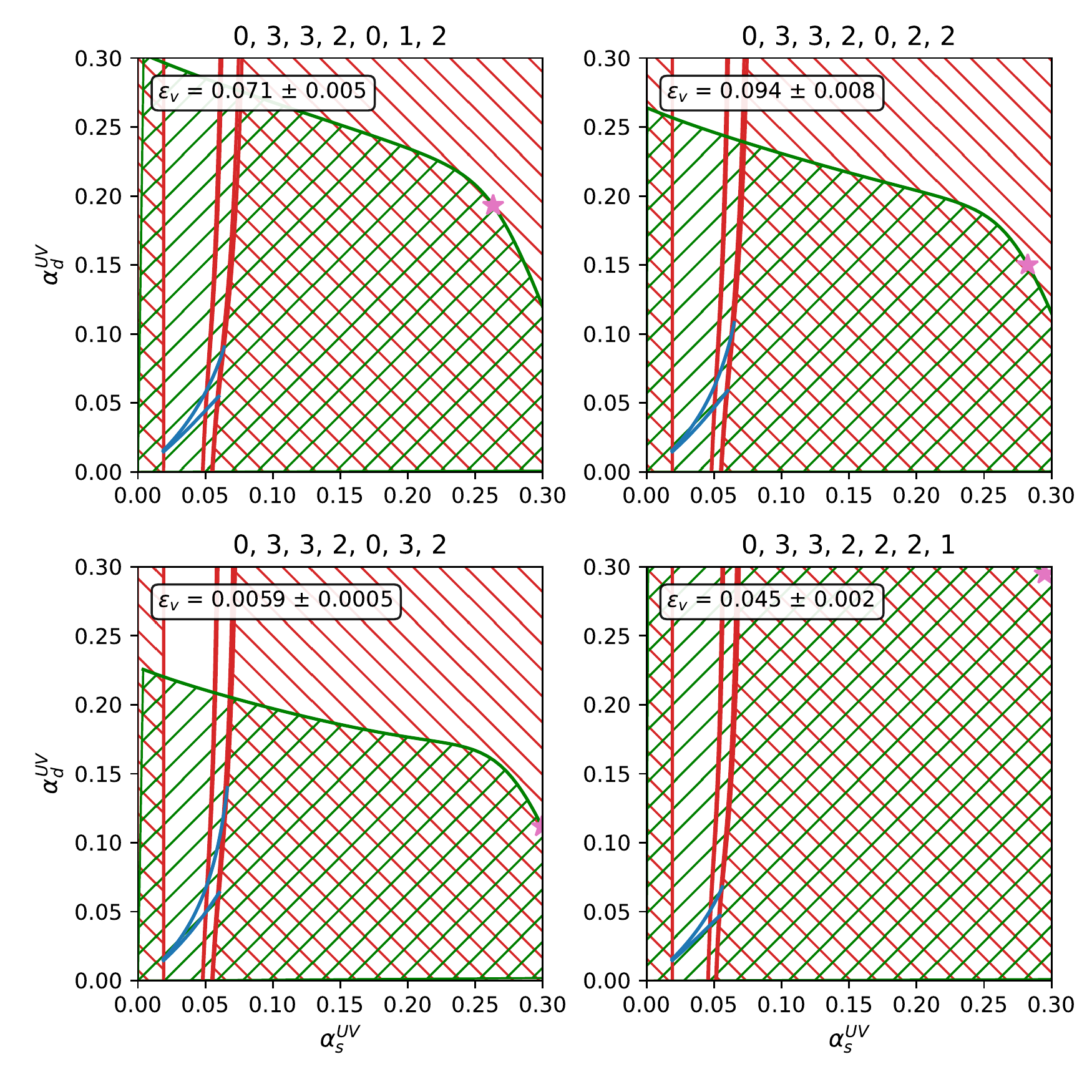}
    \caption{The results for the four models with the largest values of $\alpha_s^\star$ from the set of models with at most 3 of each new field and a perturbative IRFP. The features of each subplot are the same as those in Fig.~\ref{fig:viable_models}. Note that a larger value of $\alpha_s^\star$ correlates with a larger proportion of the parameter space that has no valid value for $M$ (the red hatched regions).}
    \label{fig:big_as*_models}
\end{figure}

We can understand easily why larger $\epsilon_v$ would correlate with smaller $\alpha_s^\star$. At the decoupling scale, $\alpha_s(\mu_0)$ must be smaller than $\alpha_s(M_Z)$ (whose value is inidicated in Fig.~\ref{fig:viable_irfps} with a black dashed line) in order to have the correct low-energy running. So, for this to occur in a decent proportion of the UV coupling parameter space, the couplings must evolve towards a fixed point where $\alpha_s^\star$ is smaller than $\alpha_s(M_Z)$. Conversely, when $\alpha_s^\star$ is greater than $\alpha_s(M_Z)$, then for many pairs of initial UV couplings with larger values of $\alpha_s^\mathrm{UV}$ there is no mass scale $M$ for which $\alpha_s$ has the correct low-energy running. This can be seen in Fig.~\ref{fig:viable_models} for the models with $\alpha_s^\star$ closest to 0.1, and is clearer in Fig.~\ref{fig:big_as*_models}, where we show the models with the four largest values of $\alpha_s^\star$.

Given that small $\alpha_s^\star$ correlates with larger $\epsilon_v$, we can also understand heuristically why $\alpha_d^\star$ should also be small in these cases. As mentioned earlier, for the visible and dark confinement scales to be of a similar order of magnitude, then $\alpha_s$ and $\alpha_d$ should generally have similar values at the decoupling scale; to achieve this for a decent proportion of the UV coupling parameter space, the couplings should thus be evolving towards a fixed point with similar values for $\alpha_s^\star$ and $\alpha_d^\star$.

The last observation we make here is about the size of the new physics scale in these models. In Fig.~\ref{fig:viable_models}, to the right of the red contour in each plot we have $M <$ 1 TeV. This would imply a plethora of new, coloured sub-TeV particles, and the model at these parameter points would be ruled out by collider constraints. So, although within our selection of 155 models we have found 12 with a decent validity fraction $\epsilon_v > 0.35$, they would be ruled out by experiment for the majority of their valid parameter space.

Now that we have identified some features of models that correlate with larger values of $\epsilon_v$, we wish to see if we can find models that have a sufficiently large new physics scale $M$ to avoid strong collider constraints in addition to having a large validity fraction.

\subsection{Searching for models with larger $M$}

To find models that have larger scales of new physics, we need to know how $M$ depends upon the selection of field multiplicities. 

Recall that $M$ is determined by the requirement that we replicate the low-energy running of $\alpha_s$: using the consistency condition of Eqn.~\ref{eqn:cons_cond_s}, we solve for the decoupling scale $\mu_0$ at which $\alpha_s(\mu_0)$ running under the $\beta$-functions of the full model matches onto $\alpha_s^\mathrm{EFT}(\mu_0)$ given by its SM running from $\alpha_s^\mathrm{EFT}(M_Z) = 0.11729$. By specifying a value of $\mu_0/M \in (0.5, 2)$ we can then solve for $M$.

The low-energy SM running of $\alpha_s^\mathrm{EFT}$ is fixed; so, to increase the value of $M$ for a given set of initial UV couplings, we need $\alpha_s$ to run more strongly under the $\beta$-functions of the full model. In general, increasing the field multiplicities will produce stronger running, as the coefficients of the terms in the $\beta$-functions increase. Since we need the coefficients of the one-loop terms to be negative in order to have an IRFP, the strongest running will occur when the coefficient of the one-loop term is negative with a small magnitude, and the coefficients of the two-loops terms are positive with a larger magnitude\footnote{Note that although the magnitude of the one-loop term will be smaller than the two-loop terms, this is due to a cancellation between the matter terms and gluon terms in the one-loop coefficient. The individual one-loop terms will still be smaller than the individual two-loop terms for perturbative $\alpha < 0.3$, as discussed in Sec.~\ref{sec:model_validity}.}. This also correlates with the IRFP couplings $\alpha_s^\star$ and $\alpha_d^\star$ having small values, as the $\beta$-functions will be zero when, roughly, the $\text{one-loop coefficient} \times \alpha^2 \sim \text{two-loop coefficient} \times \alpha^3$.

From Eqn.~\ref{eqn:beta_fn}, we see that increasing the multiplicities of the jointly-charged fields -- $n_{f_j}$ and $n_{s_j}$ -- increases the coefficients of all terms in the $\beta$-functions. Since $n_{f_j}$ can be no larger than 3 in order to keep the one-loop terms negative, we consider models with larger $n_{s_j}$ when searching for models with a larger new physics scale $M$. In particular, we look at all models with $n_{s_j} \geq 10$, where the maximum multiplicity for each field is given in Table~\ref{tab:max_mult}.

We note that in Table~\ref{tab:max_mult} there is only an upper limit on the total number of dark fermions $n_{f_d}$, and not on the specific multiplicities of the light and heavy dark fermions, $n_{f_{d,l}}$ and $n_{f_{d,h}}$; this is because the $\beta$-functions of the full model only depend on $n_{f_d}$. The new physics scale $M$ and the decoupling scale $\mu_0$ also only depend on $n_{f_d}$, as they only depend upon the SM running of $\alpha_s$ in addition to the RGE evolution using the full $\beta$-functions. So, for the purposes of searching for models with larger values of $M$, we only specify the value of $n_{f_d}$.

Using this definition of a model, there are 10,141,120 models for which $n_{s_j} > 10$, of which 120,015 have a perturbative fixed point. To have sufficiently strong running of the gauge couplings, we then look for models whose one-loop coefficients for the $\beta$-functions are negative with a small magnitude, where we take that to mean that the coefficient is between $-0.1$ and 0; there are 188 such models. For all of these models, the scale of new physics $M$ is greater than 1 TeV for at least $80\%$ of the $\alpha_s^\mathrm{UV}$-$\alpha_d^\mathrm{UV}$ parameter space; that is, the value of $M$ is large enough to avoid strict collider constraints in the majority of parameter space.

So, having found this set of models, we now wish to check which values of $\epsilon_v$ they can obtain. If we find models that have a decently large value of $\epsilon_v$, and also have sufficiently large new physics scales in the majority of parameter space, then these models could serve to naturally explain the cosmological coincidence problem.

For each of the 188 models, we have only specified $n_{f_d}$; so, for a given model we are free to choose $n_{f_{d,l}} \leq n_{f_d}$, setting $n_{f_{d,h}} = n_{f_d} - n_{f_{d,l}}$. We also ensure that $n_{f_{d,l}} > 0$ so that there is at least one light dark fermion species to confine into dark baryons. 

\begin{figure}
    \centering
    \includegraphics[width=0.45\textwidth]{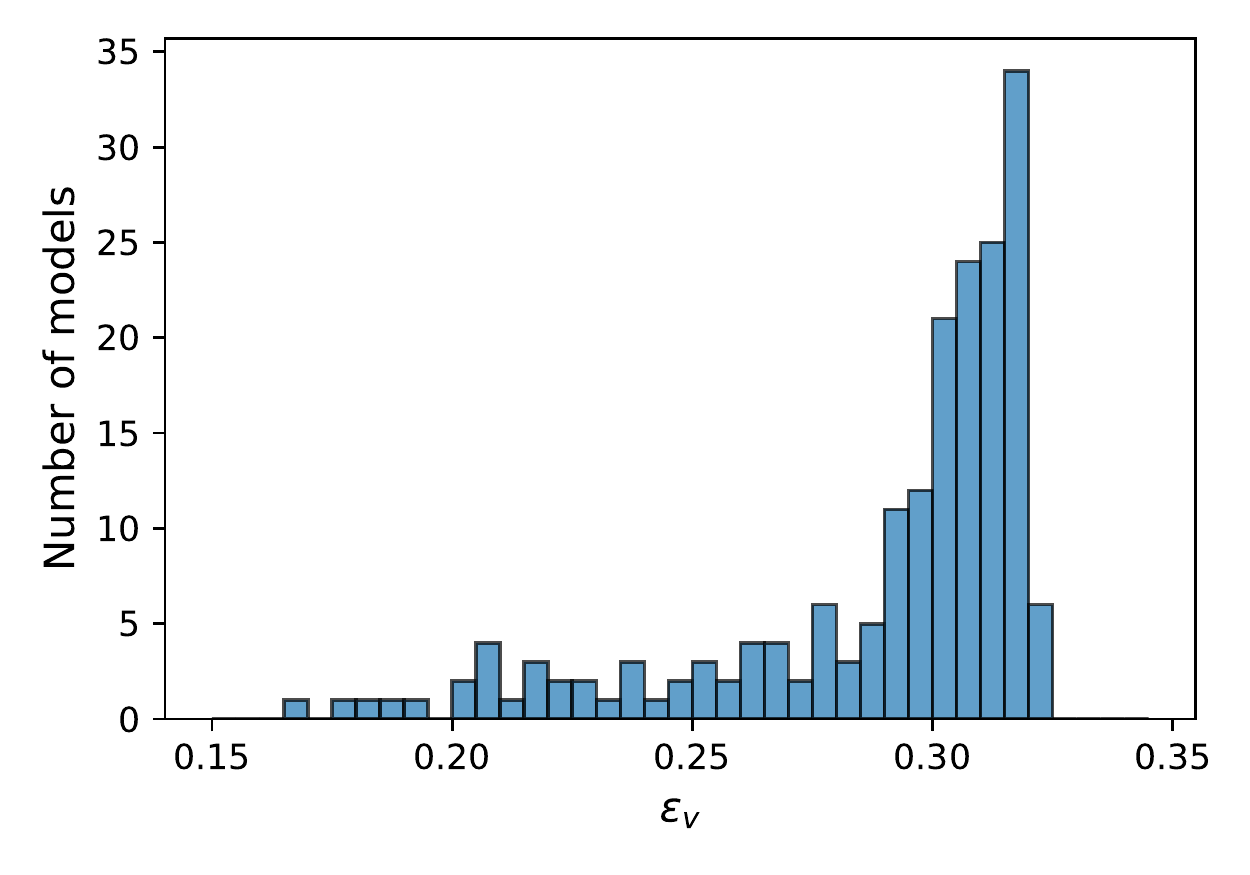}
    \caption{A histogram of the validity fractions for the set of 188 models that have 10 or more joint scalars, and one-loop coefficients for the $\beta$-functions between $-0.1$ and 0. Each of these models has a mass scale $M > 1$ TeV for at least 80\% of the UV coupling parameter space.}
    \label{fig:ev_hist_large_M}
\end{figure}

The choice of $n_{f_{d,l}}$ will change the value of the dark confinement scale $\Lambda_\mathrm{dQCD}$ for a given pair of initial UV couplings, as $\Lambda_\mathrm{dQCD}$ depends on the low-energy running of $\alpha_d$ below the decoupling scale. The $\Lambda_\mathrm{dQCD}$ contours, and thus the validity fraction $\epsilon_v$, then also depend on $n_{f_{d,l}}$. So, for each of the 188 models, we choose $n_{f_{d,l}}$ such that the value of $\epsilon_v$ is maximised. We show these values of $\epsilon_v$ in Fig.~\ref{fig:ev_hist_large_M}.

\begin{figure}
    \centering
    \includegraphics[width=0.5\textwidth]{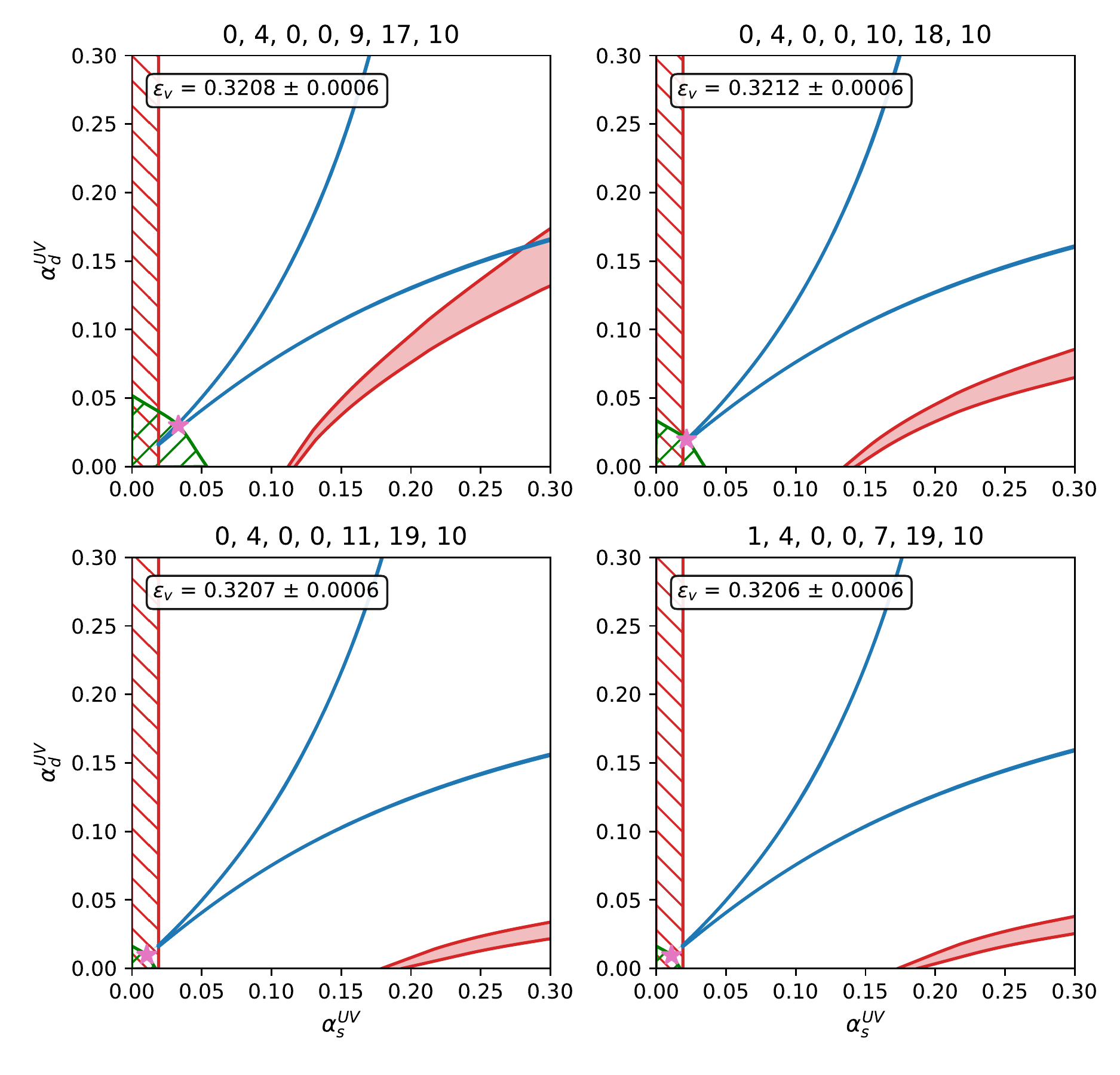}
    \caption{The results for four models with $\epsilon_v > 0.32$. These are some of the models with the largest viability fraction in the set of 188 models shown in Fig.~\ref{fig:ev_hist_large_M}. The features of each subplot are the same as those in Fig.~\ref{fig:viable_models}. Note that for each model, at least 80\% of the parameter space has $M > 1$ TeV, where $M = 1$ TeV is shown by the red contour.}
    \label{fig:large_M_models}
\end{figure}

Most of these models have values for $\epsilon_v$ above 0.3, with the largest values just over 0.32. We show 4 models with $\epsilon_v > 0.32$ in Fig.~\ref{fig:large_M_models}. While these values of $\epsilon_v$ are not quite as large as those for the best models in Sec.~\ref{sec:models_3}, a validity fraction of $\sim0.3$ still corresponds to a decent proportion of the initial UV coupling parameter space for which the dark confinement scale is on the order of the visible confinement scale. These models also have $M > 1$ TeV for every point in the viable region of parameter space, and so manage to potentially provide an explanation for the cosmological coincidence problem while avoiding stringent collider constraints.

\section{Conclusions}\label{sec:conclusions}

In this paper we reassessed the feasibility of explaining the cosmological coincidence problem by utilising infrared fixed points in a dark QCD framework. In the original work of Bai and Schwaller, the confinement scales of dark and visible QCD are related by a dynamical mechanism depending only on the value of the gauge couplings at the IRFP and the mass scale of the new particle content introduced. We extended this framework by incorporating the dependence of these results on the initial values of the gauge couplings in the UV.

To assess whether models within this framework are able to explain the cosmological coincidence problem, we needed to carefully define our notion of how `naturally’ a given model produces visible and dark confinement scales of a similar order of magnitude. With the dark confinement scale now depending on the initial UV couplings, we analysed each model by looking at the proportion of UV coupling parameter space for which one obtains a valid value for the confinement scale: denoted the `validity fraction’ $\epsilon_v$.

We began by looking at a limited set of models where we introduce at most three of each new field. Of these 12,288 models, 155 have a perturbative IRFP. The maximum validity fraction for these models is $\sim 0.4$, and there are 12 with $\epsilon_v > 0.35$. The models with larger validity fractions generally have smaller IRFPs.

While these models have a decently sized proportion of UV coupling parameter space for which the dark and visible confinement scales are related, the scale of new physics is below 1 TeV for most of this parameter space. These new coloured sub-TeV fields face strict collider constraints. To see if models in this framework can avoid these constraints, we identified a set of 188 models for which at least 80\% of the UV coupling parameter space has $M > 1$ TeV; all these models have 10 or more scalars charged under both visible and dark QCD. Of these models, most have $\epsilon_v \sim 0.3$.

In conclusion, there do exist models within this framework that can potentially provide a natural explanation for the cosmological coincidence problem. While many of these models introduce new sub-TeV coloured particles, there is a well-defined set of models that have a new physics mass scale high enough to avoid these collider constraints. In future work, these models could be candidates for more detailed model building, in order to incorporate a definite baryogenesis mechanism to generate the asymmetries in visible and dark matter, and to perform a proper phenomenological analysis.

\begin{figure*}
    \centering
    \includegraphics[width=\textwidth]{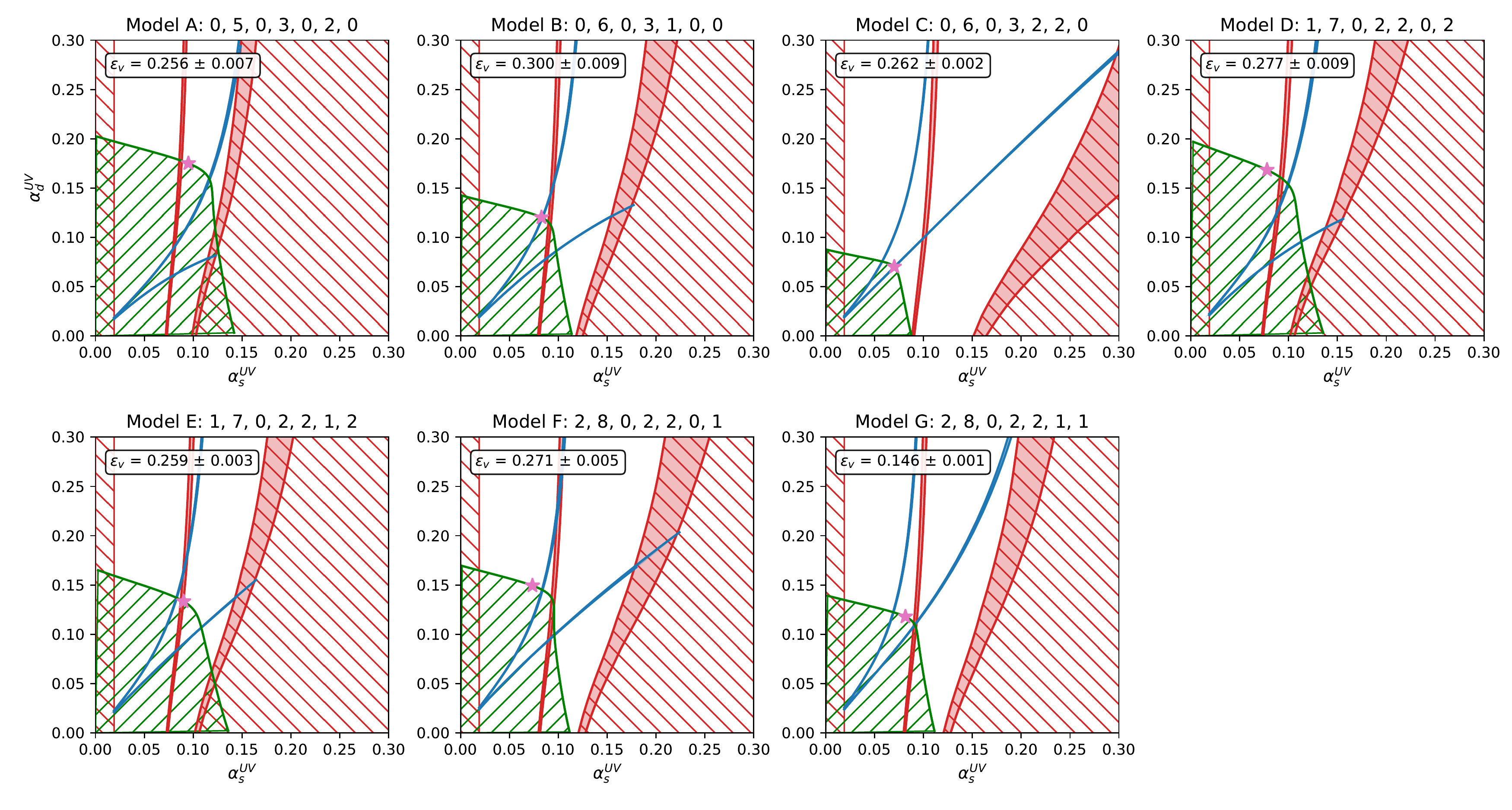}
    \caption{The results for the seven benchmark models identified in the original paper by Bai and Schwaller. The features of each subplot are the same as those in Fig.~\ref{fig:viable_models}. The results presented in the original paper correspond to our results when choosing the initial UV couplings to have their IRFP values (the point in each subplot identified by the pink star).}
    \label{fig:bs_models}
\end{figure*}

\section*{Acknowledgements}

We would like to thank Jayden Newstead and Michael Schmidt for useful conversations. ACR was supported by an Australian Government Research Training Program Scholarship. This research was supported by the Australian Government through the Australian Research Council Centre of Excellence for Dark Matter Particle Physics (CDM, CE200100008).

\bibliography{references}

\appendix

\section{The benchmark models from Bai and Schwaller}\label{sec:bs_models}

In Fig.~\ref{fig:bs_models} we plot the results for the seven benchmark models chosen by Bai and Schwaller. Table 2 of the original paper gives values of $M$ and $m_D$ (the dark baryon mass) for each model; in our analysis, these correspond to $M$ and $\Lambda_\mathrm{dQCD}$ calculated for initial UV couplings at their IRFP values, $(\alpha_s^\mathrm{UV}, \alpha_d^\mathrm{UV}) = (\alpha_s^\star, \alpha_d^\star)$, and with $m_D \approx 1.5\Lambda_\mathrm{dQCD}$. In Fig.~\ref{fig:bs_models}, these are the values for $M$ and $\Lambda_\mathrm{dQCD}$ at the point indicated by the pink star. We see that models C, E, and G have valid values of $\Lambda_\mathrm{dQCD}$ for initial UV couplings at their IRFP values; these correspond to dark baryon masses of 0.32 GeV, 3.5 GeV, and 1.2 GeV respectively. Most of the models have $\epsilon_v \sim 0.25$, with Model B having the largest validity fraction of $\epsilon_v = 0.300 \pm 0.009$. While this is a decently large validity fraction, as with the results in Sec.~\ref{sec:models_3}, most of the valid region of parameter space for these models has a new physics scale below 1 TeV (the initial UV couplings that lie to the right of the red contour).

\end{document}